\documentclass[conference]{IEEEtran}
\IEEEoverridecommandlockouts

\usepackage{cite}
\usepackage{amsmath,amssymb,amsfonts}
\usepackage{graphicx}
\usepackage{textcomp}
\usepackage{xcolor}
\usepackage[T1]{fontenc}
\usepackage[utf8]{inputenc}

\usepackage{amsmath}
\usepackage{algorithm}
\usepackage{algpseudocode}
\usepackage{hyperref}
\usepackage{listings}
\usepackage{longtable}
\usepackage[caption=false]{subfig}
\newtheorem{definition}{Definition}[section]
\usepackage{breqn}
\algrenewcommand\algorithmicindent{0.8em}

\usepackage{color}
\definecolor{deepblue}{rgb}{0,0,0.5}
\definecolor{deepred}{rgb}{0.6,0,0}
\definecolor{deepgreen}{rgb}{0,0.5,0}

\usepackage{pythonhighlight}
\usepackage{listings}
\usepackage{flushend}

\def\BibTeX{{\rm B\kern-.05em{\sc i\kern-.025em b}\kern-.08em
    T\kern-.1667em\lower.7ex\hbox{E}\kern-.125emX}}

\begin{document}

\title{
    Simulating Spreading of Multiple Interacting Processes in Complex Networks \thanks{
        This work was partially supported by the Polish National Science Centre, under Grant no. 2016/21/D/ST6/02408
    }
}

\author{
    \IEEEauthorblockN{1\textsuperscript{st} Michał Czuba}
    \IEEEauthorblockA{
        \textit{Department of Artificial Intelligence}\\
        \textit{Wrocław University of Science and Technology}\\
        Wrocław, Poland \\
        michal.czuba@pwr.edu.pl
    }\and
    \IEEEauthorblockN{2\textsuperscript{nd} Piotr Bródka}
    \IEEEauthorblockA{
        \textit{Department of Artificial Intelligence}\\
        \textit{Wrocław University of Science and Technology}\\
        Wrocław, Poland \\
        piotr.brodka@pwr.edu.pl
    }
}

\maketitle

\begin{abstract}
Investigating the interaction between spreading processes in complex networks is one of the most important challenges in network science. However, whether we would like to know how the information campaign will affect virus spreading or how the advertising campaign of the new iPhone will affect the sales of Samsung phones, we need an environment that will allow us to evaluate under what conditions our spreading campaign will be effective. \textit{Network Diffusion} is a Python package that should help do that. In this paper, we introduce its operating principle and main functionalities, including simple examples of simulations that can be performed using it.
\end{abstract}


\section{Introduction}\label{sec:intro}

For almost two years now, the resources of the entire world have been used for the fight against the "invisible" enemy. We spent countless hours and trillions of dollars to stop the COVID-19 pandemic, and still, millions of people have died from the virus or the collapse of healthcare systems around the world. One of the challenging tasks is understanding how different countermeasures, such as information campaigns, lockdowns, or vaccinations, affect the progression of the virus to choose those that will save both human lives and the economy.

In this paper, we would like to present the \textit{Network Diffusion} package~\cite{czuba2021zenodo}that allows simulating the spread of many coexisting processes in single and multilayer networks. An example of such processes would be, of course, coronavirus spread, interacting with an information campaign and vaccinations, but also the interaction between two advertising campaigns (e.g. iPhone vs Samsung or Cannon vs Nikon), two political campaigns (e.g. Biden vs Trump), diseases (e.g., AIDS and Tuberculosis), to name just a few~\cite{brodka2020interacting}. 
However, before we can present the \textit{Network Diffusion} package, we would like to introduce a few key concepts from network science~\cite{barabasi}.

\subsection{Complex networks}\label{subsec:complex_nets}

Complex networks are the backbone of all complex systems~\cite{barabasi} starting from the nervous system of living organisms~\cite{stark2006biogrid}, through our infrastructure systems (electric grid, water pipes, airports, etc.)~\cite{de2014navigability, watts1998collective}, and ending with relationships between people~\cite{kazienko2011multidimensional}. Complex networks can be represented and analysed in multiple ways, e.g., adjacency matrix~\cite{barabasi} or property vectors~\cite{brodka2018quantifying}, however, the most common approach is to represent a complex network as a graph~\cite{barabasi} or a set of graphs (for multilayer networks)~\cite{kivela2014multilayer, magnani2011ml}. The multilayer network, according to~\cite{kivela2014multilayer}, is defined as quadruple $M = (N,L,V,E)$; where: $N$ is a set of actors; $L$ is a set of layers; $V$ is a set of nodes, $V \subseteq N \times L$; $E$ is a set of edges $(v_1, v_2): v_1, v_2 \in V$, and if $v_1=(n_1, l_1)$ and $v_2=(n_2, l_2) \in E$ then $l_1=l_2$.

\subsection{Spreading phenomena}
Spreading phenomenon covers a wide spectrum of processes, starting from information diffusion~\cite{guille2013information}, through virus~\cite{kermack1927contribution}, opinion~\cite{holley1975ergodic}, innovation~\cite{bass1969new} spreading and ending with the spread of influence~\cite{kempe2003maximizing}. Fortunately, all these processes are similar if we look at their high-level components~\cite{barabasi} (i) \textit{What} phenomenon is spreading (e.g. information, opinion, virus), (ii) \textit{How/Who} it is spreading (e.g. network nodes, actors, agents), and (iii) \textit{Where} it is spreading (i.e. network type). All these elements carry a different context to the simulation results, but, at the same time, they can be imagined as "containers" for algorithmic structures, which will be introduced later. An example of those three components can be a viral video that spreads on a social networking site. Here, \textit{what} is the video or digital content, \textit{how/who} is a link to the website or a post on social media, and \textit{where} is a social network that is the foundation of this particular social networking site.

\subsubsection{Epidemic modelling framework}\label{subsubsec:epidemic_framework}
The initial models developed by epidemiologists are based on two assumptions~\cite{barabasi}: compartmentalisation and homogeneous mixing. The first condition says that the state of each individual is discrete. This means (continuing with epidemiological analogies) that the relation of a given node to a disease can be described by $n$ states, e.g. $healthy$, $sick$. The second assumption concerns the ability of the nodes to change their states. Each individual can interact with all nodes in a given time unit. Using that hypothesis, we can build many various spreading models like SI (Suspected-Infected), SIS (Suspected-Infected-Suspected), SIR (Suspected-Infected-Recovered), SIRS (Suspected-Infected-Recovered-Suspected), and so on. Epidemic models can be used to simulate other spreading processes. The most common example would be information spreading, where we have models like UAU~\cite{zang2018effects} (Unaware-Aware-Unaware) or UAF~\cite{scata2016impact} (Unaware-Aware-Forgot), which are based on SIS and SIR models, respectively. 

\subsubsection{Independent cascade model}\label{subsec:ICM}
Independent cascade model (ICM)~\cite{kempe2003maximizing} has a much different mechanism and is widely used to simulate influence diffusion. Its principles reject the homogeneous mixing assumption because of the way the phenomenon propagates - it cannot be captured holistically for a given network. Here, the spreading takes the form of a cascade. Each newly activated node has one chance to activate its neighbours in each step. It is based on the likelihood of activation (aka. propagation probability) stored at the edge connecting them~\cite{diffusion}.

\subsubsection{Linear threshold model}\label{subsec:LTM}
This model is based on the concept of the activation threshold that is defined for each node of the network~\cite{kempe2003maximizing}. It determines a minimum value of the influence of its neighbours to change its state. In other words, the minimum value above must be the sum of the intensity of the connections from the neighbouring activated nodes to result in activation~\cite{diffusion}. When comparing ICM and Linear Threshold Model (LTM), one can say that the first one is based on the \textit{push} mechanism, i.e., a node pushes its influence to its neighbours. In contrast, LTM is based on the \textit{pull} mechanism, i.e. the node pulls the influence from its neighbours.

\subsection{Interacting spreading processes in networks}\label{subsect:multilateral}
When analysing how various phenomena in networks propagate, we have to ask the following questions: \textit{what is being propagated?} and \textit{where is it being propagated?}. As an answer, we get four general possibilities: (1) a single process in a single network, (2) a single process in a multilayer network, (3) multiple processes in a single network, and (4) multiple processes in a multilayer network.

The last two options bring the possibility of spreading multiple processes that can interact with each other. Thus, in addition to modelling each phenomenon, we also need to model the interactions between them, i.e., how they influence each other (and that is exactly a problem that \textit{Network Diffusion} solves). Below, we briefly describe the four possible variants of interactions.

Firstly, we can distinguish supporting processes (about 11\% of research in the domain~\cite{brodka2020interacting}) that mutually support each other by increasing coverage and velocity. Here, a good illustration is that chronic diseases (such as asthma) are a catalyst for contracting COVID-19.

The next genres are competing processes (about 36\% publications~\cite{brodka2020interacting}). In this case, one process causes suppression of the propagation of the other. A good example of this is the presidential election: if the number of people influenced by candidate X increases, the number of people influenced by candidate Y has to decrease, and at the end of the day, only one candidate can win.

The third case, the mixed approach, covers about 46\% of the publications~\cite{brodka2020interacting}. It considers instances where one process supports the second, but the second competes with the first. An interaction between disease and awareness can be a good example. People who get sick become aware of a disease, so the more people get sick, the more people will be aware. On the other hand, people aware of the pandemic will take preventive actions to limit its spreading.

The last case is when processes do not interact with each other. It appears only in 7\% of the papers in the domain~\cite{brodka2020interacting}.

\section{Similar software packages}\label{sec:competition}

To gain valuable recognition of similar solutions to \textit{Network Diffusion}, an appropriate tool review methodology was adopted. For this, the general state of the software available in the field was taken into account. We started from a simple reconnaissance using a search engine and focused more on the tools available for the Python language; however, it was not a strict condition. Our final goal was to see a general cross-section of state-of-the-art. As a result, information on more than twenty different tools was obtained. We analysed them and divided them into two groups: \textit{packages dedicated for network spreading simulations} and \textit{general complex network analysis software}.

\subsection{Software for spreading processes simulations}\label{subsec:close_competition}

\begin{table*}[ht]
    \renewcommand{\arraystretch}{1.3}
    \caption[]{Tools with corresponding functionalities to \textit{Network Diffusion}.}
    \label{tab:similar_tools}
    \centering
    \begin{tabular}{|l|p{0.07\textwidth}|l|p{0.61\textwidth}|l|}
    \hline \textbf{Name} & \textbf{Type} & \textbf{Environment} & \textbf{Functionalities} & \textbf{Access} \\ \hline \hline
    GLEaMviz & app. & desktop & single model propagation, high freedom of model definition, visualisation of the experiment~\cite{gleam} & free \\ \hline
    NDLIB & library & Python & single model propagation, high freedom of model definition, visualisation of the experiment~\cite{ndlib} & free \\ \hline
    SimInf & library & R & propagation of custom model, visualisation~\cite{siminf} & free \\ \hline
    Sispread & app. & console & SI, SIS or SIR model propagation, returns numerical data~\cite{sispread} & free \\ \hline
    STEM & Eclipse plugin & desktop & single model propagation, visualisation, workflow like finite element method (FEM) environments~\cite{stem} & free \\ \hline
    \end{tabular}
\end{table*}

\subsubsection{GLEaMviz}

The first application that has functionalities corresponding to the designed software is a GLEaMviz. It works with real data, population density, and migration around the world, combined with stochastic models of disease propagation. As a result, it provides a sophisticated simulation environment. Due to the large scale of the experiments (the whole world), a single node is a population of a given size (defined by the user). A very interesting feature is the manual definition of the epidemiological model. GLEaMviz makes this possible by manipulating the compartments (understood in the same way as in sec.~\ref{subsubsec:epidemic_framework}). Allowable transitions between them are also fully definable. The user can also select the geographical start of the disease, the initial percentages of individuals belonging to a given compartment, its duration, etc. There is also an option to generate various visualisations at the end of the experiment. Despite the interesting functionalities mentioned above, GLEaMviz has a rather large disadvantage: it only allows the propagation of one process at a time~\cite{gleam}.

\subsubsection{NDLIB}\label{ndlib}

Network Diffusion LIBrary is a Python package based on the NetworkX library. It allows performing simulations with many predefined epidemiological models (such as: SIS, SIR, SEIR, etc.), influence group (LTM, ICM, Profile, etc.), opinion group (Voter, Sznajd, etc.), and even dynamics (models with the capacity to change the topology of network). Moreover, the user can create its own customised models. Results visualisation is also possible via Matplotlib or Bokeh with the flexibility to append a custom graphical engine. NDLIB also has some interesting run-time features. First of all, it includes an option to perform a "multi-execution" of the simulation by parallel computing. As this kind of experiment is generally stochastic, this feature gives a chance to see the general behaviour of the observed phenomena. It also enables running the simulation on a server (as well as locally). For users unfamiliar with Python, the authors created NDQL, a query language (based on SQL syntax) that supports elementary NDLIB commands. For those who cannot programme, they also provided a "visualisation framework" to play with some of the models implemented with a GUI-based tool~\cite{ndlib}. NDLIB is a very useful library with many features. However, it does not directly support experiments where many processes interact together.

\subsubsection{SimInf}

The next software is SimInf, a process diffusion package for R. Its API has been designed in a very interesting way. This tool allows a user to define his own models by parsing the appropriate string. Interestingly, no network is needed to run the simulation, as the algorithms work on the assumption of homogeneous mixing. After a successful simulation, it is possible to display a summary graph. Nevertheless, we must note that no support for operations on real networks and the lack of an option to define multiprocess experiments~\cite{siminf} are distinct shortcomings.

\subsubsection{Sispread}

Sispread is a simple application implemented in C. It is a console tool without a graphical interface. It focuses on three models (SI, SIS, SIR) with the possibility of deep analysis of the experiments performed. This tool supports very basic IO operations - the user can upload a custom network that meets certain requirements (without the option to manipulate it). As a result of the experiment, numerical data are returned for further analysis~\cite{sispread}.

\subsubsection{STEM}

Another advanced system is STEM. As an extension of the well-known Eclipse IDE, it uses its graphical layout and the general philosophy of user interaction. It is reminiscent of FEM systems in the way it works. Just like them, it requires the user to specify a medium (equivalent to a model of a physical object) where the simulation is performed with its discretisation level (i.e. whether the node is a municipality or an entire county). The next step is to attach an appropriate solver propagation model and set the starting parameters of the experiment. Once this is done, it is possible to visualise the spread progress. As in other programmes, it does not support multiple spreading processes~\cite{stem}.

\subsection{Software for complex network analysis}\label{sec:other_tools}

Due to the large number of tools found (see tab.~\ref{other_tools}), we describe below only the most interesting of them.

\begin{table*}
    \renewcommand{\arraystretch}{1.3}
    \caption[]{Software for complex network analysis.}
    \label{other_tools}
    \centering
    \begin{tabular}{|l|l|l|l|l|}
    \hline
    \textbf{Name} & \textbf{Type} & \textbf{Environment} & \textbf{Functionalities} & \textbf{Access} \\ \hline \hline
    AllegroGraph & DBMS & desktop & graph database management system, visualisation, static analysis of networks~\cite{allegrograph} & free \\ \hline
    Arcgis & app. & desktop, web & static analysis of multilayer networks, visualisation, IO~\cite{arcgis} & free \\ \hline
    Cytoscape & app. & desktop & static analysis of networks, visualisation~\cite{cytoscape} & free \\ \hline
    Gephi & app. & desktop & static analysis of networks, visualisation~\cite{gephi} & free \\ \hline
    Giraph & library & Java & static analysis of networks~\cite{giraph} & free \\ \hline
    Graphviz & app. & console & visualisation~\cite{graphviz} & free \\ \hline
    iGraph & library & Python, R, C, M & static analysis of multilayer networks, visualisation, IO~\cite{igraph} & free \\ \hline
    Multinet & package & R Python & manipulation of multilayer networks, static analysis, visualisation IO~\cite{multinet} & free \\ \hline
    Muxviz & app. & desktop & manipulation of multilayer networks, static analysis, static and dynamic visualisation ~\cite{muxviz} & free\\ \hline
    Neo4j & app. & desktop & graph database management system, visualisation, static analysis of networks~\cite{neo4j} & paid \\ \hline
    NetMiner & app. & desktop & static analysis of networks, visualisation, IO, Python-like scripting~\cite{netminer} & paid \\ \hline
    Network & library & R & static analysis of multilayer networks, IO~\cite{packagenetwork} & free \\ \hline
    NetworkX & library & Python & static analysis of networks, visualisation, IO~\cite{networkx} & free \\ \hline
    Nodexl & plugin & Excel & static analysis of networks, visualisation, IO~\cite{nodexl} & free, paid \\ \hline
    Pymnet & library & Python & static analysis of multilayer networks, visualisation, IO~\cite{pymnet} & free \\ \hline
    Tulip & app. & desktop & static analysis of networks, visualisation, IO, Python-like scripting~\cite{tulip} & free \\ \hline
    yEd & app. & desktop & static analysis of networks, visualisation, IO~\cite{yed} & free, paid \\ \hline
    \end{tabular}
\end{table*}

\subsubsection{Gephi}

Gephi is an application for visualising networks in many aspects, with extensive graphical capabilities. It allows the user to perform exploratory data analysis (EDA) in real-time on basic structures and connections between objects. Gephi is also designed for social network analysis - it has an easy way to map different portals (Facebook, Twitter, etc.) and Small World networks with each other. It can also visually represent biological data and serve as a tool for creating scientific posters. Additionally, within Gephi, the basic metrics for network analysis have also been implemented. The package works with many file types and supports integration with plugins written by external developers. In terms of non-functional aspects, it has an interface that does not require programming skills~\cite{gephi}.

\subsubsection{iGraph}

iGraph is a library designed for several programming languages (i.e. C, R, Python, and Mathematica). The three main paradigms of iGraph are: easy handling of graph algorithms, fast handling of large networks, and enabling interaction with high-level languages like R. In many aspects, this tool is similar to pure NetworkX, but it is also able to run simple epidemic models like SIR. For clarity, it should be added that this functionality is rather residual~\cite{igraph}.

\subsubsection{NetMiner}

NetMiner is a comprehensive environment for network analysis and visualisation. Unlike the tools described above, it is not free software. Among the functionalities that distinguish it from the others, we can list for sure the possibility of EDA (e.g., analysis of written text and, on this basis, creation of semantic network analysis), clustering, classification, etc. NetMiner also has extensive visualisation capabilities, both static and dynamic, and, what is interesting, three-dimensional. Another important feature of this tool is the presence of a Python-based console. It can even be upgraded by adding external packages. Moreover, NetMiner allows the compilation of written programmes, which is a big advantage of it in terms of securing the code against third parties~\cite{netminer}.

\subsubsection{NetworkX}

NetworkX is one of the most important packages among Python data science libraries. Its capabilities are large and mainly concern static network analysis. NetworkX also has visualisation functionality, but in its official documentation, it recommends the use of Cytoscape, Gephi, and Graphviz to perform any visualisation~\cite{networkx}. Furthermore, because of its integration with the Python language, the long stability of this library can be assumed. NetworkX has a large number of algorithms and metrics that are useful for network analysis, which makes it virtually unrivalled among other Python programming tools. Moreover, this library is a base for several tools listed in this review.

\subsection{Summary}

Having the analysis of the existing solutions done, it is easy to see that they can be divided into several types based on the main functionality: (1) software for static analysis of networks, (2) software for visualising networks, and (3) software for simulation of process propagation in networks. We can also distinguish the following groups based on the type of the tool: (1) web applications, (2) locally executed applications, (3) libraries for languages popular in data science, and (4) others (database management systems, plugins for other applications, etc.). However, it is worth noting that, after careful reconnaissance, virtually no simulation environment was found for the simultaneous propagation of multiple processes.

\section{Operating principle}\label{sec:operating_principle}

The core functionality of the software requires adequate attention, especially since, as demonstrated in sec.~\ref{sec:competition}, no solution was found that covers the problem of spreading multiple processes in networks. Hence, before the start of designing the software, we had to address and answer two questions: (1) how to run any number of processes during a single experiment, and (2) how to determine (in a general way) the interactions between processes? To make further deliberations easier to understand, we defined a domain dictionary with all important terms (see tab.~\ref{domain_dict})

\begin{table*}[ht!]
    \renewcommand{\arraystretch}{1.3}
    \caption{Domain dictionary for terms introduced to design the \textit{Network Diffusion} package.}
    \label{domain_dict}
    \centering
    \begin{tabular}{|p{0.07\textwidth}|p{0.18\textwidth}|p{0.68\textwidth}|}
    \hline
    \textbf{Term} & \textbf{Symbol} & \textbf{Description} \\ \hline \hline
    Epoch & $\varepsilon_{i} \in \epsilon : \{ 1, ..., \varepsilon\} $ & A discrete-time unit of the Experiment when all Nodes of the Network have updated their State against each of the Processes. \\ \hline
    Experiment & $X = (\mathfrak{M}, M, \epsilon)$ & A complete run of the Propagation model on the Network for a defined number of Epochs. \\ \hline
    Network & $M = (N,L,V,E)$ & A multilayer network (as introduced in sec.~\ref{subsec:complex_nets}). \\ \hline
    Process & $T=[t_{1}, t_{2}, ..., t_{k}]$ & Singular phenomena affecting the Network, described by an ordered set of States. Each Process must consist of at least two States and, in the package, is identified with the particular Network Layer. \\ \hline
    Propagation model & $\mathfrak{M} = (T_{1} \times \hdots \times T_{n}, W) $ & A definition of how Processes should interact with each other, i.e. a set of all possible global States of the Nodes and a set of Weights under which Transitions are allowed during Experiment. \\ \hline
    State & $t_{ajx}:$ \textcolor{white}{...................................} $t_{a} \in T_{1}; t_{j} \in T_{2}; t_{x} \in T_{3}$ & Categorical value denoting a relationship of the Node against a particular Process (a local State - here e.g. $t_{a}$) or a set of interacting Processes (a global State - here $t_{ajx}$).\\ \hline
    Transition & $t_{aj} \xrightarrow{} t_{ak}:$ \textcolor{white}{.................} $ t_{a} \in T_{1}; \hspace{1em} t_{j}, t_{k} \in T_{2}$ & Possible change of State of the Node at potential denoted by the Weight. Transitions are allowed between neighbouring States, e.g. they must differ only by one State. \\ \hline
    Weight & $w_{t_{j}}^{t_{k}} \in W$ & Pseudo-probability of particular Transition occurrence (here: $t_j \xrightarrow{} t_k$), valued by a number in range $[0, 1]$. \\ \hline
    \end{tabular}
\end{table*}

\subsection{Generalisation of single processes modelling}\label{subsec:single_process}

According to the concepts described in sec.~\ref{sec:intro}, we can notice a similarity between many discrete processes. Moreover, due to the numerical nature of the simulations carried out, it is not possible to hold the assumption of homogeneous mixing. Taking these into account, it is easy to conclude that models like SIS, SIR, SIRS, etc. can be reduced to specific cases of a model with $T=[t_{1}, t_{2}, ..., t_{k}]$ possible states, where each of the network nodes can be in one of them at a given time step. Moreover, the transition from one state to another takes place under a certain weight, which can be called pseudo-probability. Let us further note that, e.g. in the SIR model, it is impossible to transit directly from $S$ to $R$. Thus, a state change is only possible between neighbouring states, i.e. from $t_{k}$ one can only move to $t_{k-1}$ and $t_{k+1}$. 

\begin{definition}\label{def:process}
    It follows that every model $\mathfrak{M}$ belonging to the mentioned group can be described for the purposes of the designed system by: vector of states $T=[t_{1}, t_{2}, ..., t_{k}]$ and vector of transition weights between neighbouring states $W=[w^{2}_{1}, w^{3}_{2}, ..., w^{k}_{k-1}]$ (where $w_{n}^{m}$ means a weight of transition $n\xrightarrow{}m$); which can be combined in a matrix: \\
	\begin{center}
		\begin{tabular}{llllll}
            & $t_{1}$ & $t_{2}$ & $t_{3}$ & $\hdots$ & $t_{k}$ \\ \cline{2-4} \cline{6-6} 
            \multicolumn{1}{l|}{$t_{1}$} & \multicolumn{1}{l|}{} & \multicolumn{1}{l|}{$w^{2}_{1}$} & \multicolumn{1}{l|}{} & \multicolumn{1}{l}{} & \multicolumn{1}{|l|}{$w^{k}_{1}$} \\  \cline{2-4} \cline{6-6} 
            \multicolumn{1}{l|}{$t_{2}$} & \multicolumn{1}{l|}{$w^{1}_{2}$} & \multicolumn{1}{l|}{} & \multicolumn{1}{l|}{$w^{3}_{2}$} & \multicolumn{1}{l}{} & \multicolumn{1}{|l|}{} \\ \cline{2-4} \cline{6-6} 
            \multicolumn{1}{l|}{$t_{3}$} & \multicolumn{1}{l|}{} & \multicolumn{1}{l|}{$w^{2}_{3}$} & \multicolumn{1}{l|}{} & \multicolumn{1}{l}{} & \multicolumn{1}{|l|}{} \\ \cline{2-4} \cline{6-6} 
            \multicolumn{1}{l}{$\vdots$} & & & & \multicolumn{1}{l}{$\ddots$} & \\ \cline{2-4} \cline{6-6} 
            \multicolumn{1}{l|}{$t_{k}$} & \multicolumn{1}{l|}{$w^{1}_{k}$} & \multicolumn{1}{l|}{} & \multicolumn{1}{l|}{} & \multicolumn{1}{l|}{} & \multicolumn{1}{l|}{} \\ \cline{2-4} \cline{6-6} 
		\end{tabular}
	\end{center}
\end{definition}
\vspace{10pt}

For example, a SIRS model with coefficients: $\beta$ (individual's infection rate), $\mu$ (individual's recovery rate), and $\xi$ (loss of immunity) can be expressed as a vector of states: $T=[s, i, r]$; vector of transition weights between neighbouring states: $W=[\beta, \mu, \xi]$, and process matrix: 
\begin{center}
    \begin{tabular}{llll}
    	& $s$                   & $i$                          & $r$                        \\ \cline{2-4} 
    	\multicolumn{1}{l|}{$s$} & \multicolumn{1}{l|}{} & \multicolumn{1}{l|}{$\beta$} & \multicolumn{1}{l|}{}      \\ \cline{2-4} 
    	\multicolumn{1}{l|}{$i$} & \multicolumn{1}{l|}{} & \multicolumn{1}{l|}{}        & \multicolumn{1}{l|}{$\mu$} \\ \cline{2-4} 
    	\multicolumn{1}{l|}{$r$} & \multicolumn{1}{l|}{$\xi$} & \multicolumn{1}{l|}{}        & \multicolumn{1}{l|}{}      \\ \cline{2-4} 
    \end{tabular}
\end{center}

\subsection{Example of two processes model}\label{subsec:example_two_proc}

The way we defined the coexistence of several processes looks similar, but it is scaled into additional dimensions. As we know from sec.~\ref{subsect:multilateral}, processes interactions can be supporting, competing, mixed, or independent. Let us consider a two-process model $\mathfrak{M} = (T_{1} \times T_{2}, W)$ representing propagation of SIR-like disease ($T_{2}$) and UA-like vaccine against it ($T_{1}$) - a classic example of competing processes. Thus, we have:

\begin{itemize}
    \item $T_{1} = [u, v]$; $u$ - unvaccinated individual, $v$ - vaccinated individual
    \item $T_{2} = [s, i, r]$; $s$ - suspected individual (i.e. not yet affected by illness), $i$ - infected individual, $r$ - recovered individual
\end{itemize}

\begin{figure}
    \centering
    \includegraphics[width=0.68\columnwidth]{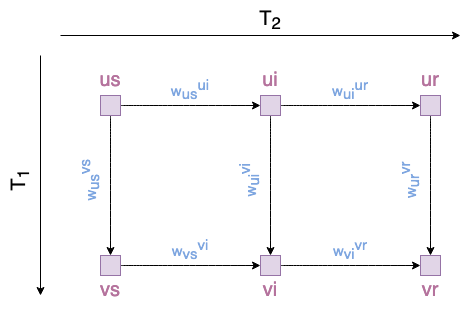}
    \caption[]{Example of two process model
    $\mathfrak{M} = (T_{1} \times T_{2}, W)$;
    $T_{1} = [u, v]$; $T_{2} = [s, i, r]$; $W = [w_{us}^{ui}, w_{ui}^{ur}, w_{vs}^{vi}, w_{vi}^{vr}, w_{us}^{vs}, w_{ui}^{vi} w_{ur}^{vr}]$.}
    \label{fig:sisir_example}
\end{figure}

Both processes affect each node of the network. Therefore, we can describe all possible states in the model with the following set (a Cartesian product - $T_{1} \times T_{2}$): $\{us, ui, ur, vs, vi, vr\}$. Moreover, under the assumption that at each simulation step, it is possible only to change the state of a node in the dimension of the singular process, we can list all available transitions (see fig.~\ref{fig:sisir_example} for visual presentation):

\begin{itemize}
    \item for $T_{1}$ (vaccination process); for $s$ constant: $us \xrightarrow{} vs$, for $i$ constant: $ui \xrightarrow{} vi$, for $r$ constant: $ur \xrightarrow{} vr$, 
    \item for $T_{2}$ (disease process); for $u$ constant: $us\xrightarrow{} ui \xrightarrow{} ur$, for $v$ constant: $vs \xrightarrow{} vi \xrightarrow{} vr$ 
\end{itemize}

Please note, that if we were to assign appropriate weights to particular transitions that reflect the probability of a state change in one process given the others, we could obtain the effect of reinforcing or suppressing between processes. For example, $w_{us}^{ui} > w_{vs}^{vi}$ means that it is easier for unvaccinated individuals to get infected than for those who are vaccinated.

Moreover, with a more detailed analysis of fig.~\ref{fig:sisir_example}, we can see the advantages of the assumption that transitions are allowed only in the direction determined by the singular process. For our example, it means that changes of node's state can occur only in the $T_{1}$ axis (e.g. $us \xrightarrow{} vs$) or the $T_{2}$ axis (e.g. $us \xrightarrow{} ui$), but never "diagonally". This eliminates obviously incorrect transitions such as $ui \xrightarrow{} vr$, which denotes a change of an unvaccinated and suspected individual to vaccinated and recovered one in one time step without even getting infected.

This assumption is also beneficial for theoretically possible transitions, where, at the same time step, not one but two or more processes change their states. First of all, it greatly simplifies defining the impact of one process on another. Furthermore, looking ahead, it greatly simplified the implementation of \textit{Network Diffusion}. Let us consider all theoretically possible transitions $us \xrightarrow{} vi$, (unvaccinated, suspected) $\xrightarrow{}$ (vaccinated, infected). Let us also assume that the fact of being vaccinated significantly reduces the probability of getting infected. Then we have three ways to get from $us$ to $vi$:

\begin{enumerate}
    \item (unvaccinated, suspected) $\xrightarrow{w_{us}^{vs}}$ (vaccinated, suspected) $\xrightarrow{w_{vs}^{vi}}$ (vaccinated, infected)
    \item (unvaccinated, suspected) $\xrightarrow{w_{us}^{ui}}$ (unvaccinated, infected) $\xrightarrow{w_{ui}^{vi}}$ (vaccinated, infected) 
    \item (unvaccinated, suspected) $\xrightarrow{w_{us}^{vi}}$ (vaccinated, infected) 
\end{enumerate}

In such a context, it is easy to see that setting $w_{vs}^{vi} << w_{us}^{ui}$ models the real-world scenario that vaccination significantly reduces the probability of infection. The other weights are also intuitive to interpret, with one exception. The value of ${w_{us}^{vi}}$, makes impossible to deduce the relation between $T_{1}$ and $T_{2}$.

\subsection{Generalisation of modelling multiple processes}\label{subsec:generalisaiton_of_model}

Using the example presented in the sec.~\ref{subsec:example_two_proc}, it is possible to define a general scheme of the multiprocess model that is used in the \textit{Network Diffusion}:

\begin{definition}
    The mutual impact between $n$ processes meeting def.~\ref{def:process}, can be represented by a $n$-dimensional orthogonal grid (spanned by points formed by the Cartesian product of the sets of states of each process), in which the connections between points exist only in directions along the axes denoting processes (i.e. transitions are allowed only between points differing by a single coordinate).
    \label{def:metamodel}
\end{definition}

This approach sufficiently answers the two questions posed at the beginning of this section; i.e. it allows us to define any number of processes within a single experiment and to determine their influence on each other clearly. 

\begin{figure}[ht]
    \centering
    \subfloat[]{\includegraphics[width=0.68\columnwidth]{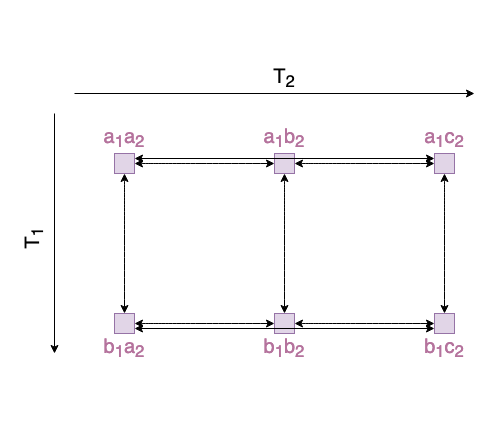}\label{fig:ab_abcimg}}\quad
    \subfloat[]{\includegraphics[width=0.68\columnwidth]{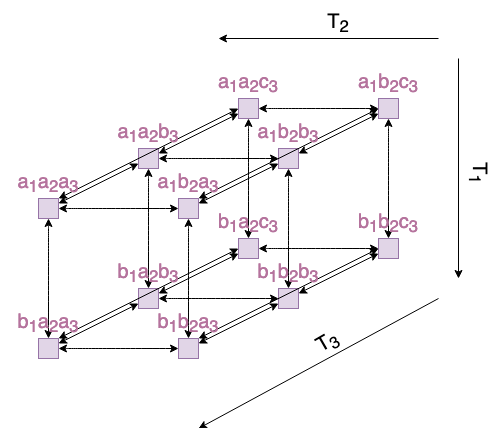}\label{fig:ab_ab_abcimg}}
    \caption[]{
        General examples of multiprocess models (top) $\mathfrak{M} = (T_{1} \times T_{2}, W)$, $T_{1} = [a_{1}, b_{1}]$, $T_{2} = [a_{2}, b_{2}, c_{2}]$; (bottom) $\mathfrak{M} = (T_{1} \times T_{2} \times T_{3}, W)$, $T_{1} = [a_{1}, b_{1}]$; $T_{2} = [a_{2}, b_{2}]$; $T_{3} = [a_{3}, b_{3}, c_{3}]$; For readability of diagrams vectors of weights $W$ are not depicted.
    }
\end{figure}

\subsection{General simulation framework}\label{subsec:sim_framework}

The concepts presented allow us to assume that there is only one process running in a given layer of the network. This means that the network should have as many layers as many processes are simulated. As a result, with the multilayer structure, each node will be described in the context of each of the defined processes. However, if the layers do not have an identical set of vertices, it will mean that a given node is indifferent to the process in whose layer it does not exist (e.g., a person who does not use social media and cannot be affected by information spreading via Twitter). 

Continuing this line of reasoning, one can arrive at a general simulation scheme. For the convenience of the reader, it is presented as alg.~\ref{simalgo}. This allows simulations to be run on custom, discrete models with a virtually infinite number of processes and internal states.

\begin{algorithm}[ht!]
    \caption{Schema of the simulation in \textit{Network Diffusion}.} \label{simalgo}
    \begin{algorithmic}[1]
        
        \Require $\mathfrak{M}$
        \begin{equation}
            \mathfrak{M}=(T_{1} \times \hdots \times T_{k}, W)
        \end{equation}
        
        \Require $M$
            {\small \begin{gather}
                M=(N,L,V,E): \\
                \nonumber \bigwedge_{i \in \{1, ..., k\}} L_{i} \xrightarrow{} T_{i}; \\ 
                \nonumber \bigwedge_{n \in N}
                n \xrightarrow{} \dot{\tau_{n}} = \{ \tau_{n}^{1}, ..., \tau_{n}^{k} \} \equiv \{ t_{1} \in T_{1} \vee \emptyset, ..., t_{k} \in T_{k}  \vee \emptyset \};
            \end{gather} }
       
        \Require $\epsilon$
        \begin{equation}
            \epsilon = \{ 1, ..., \varepsilon\}
        \end{equation}
        
         \Procedure{$X$}{$\mathfrak{M}, M, \epsilon$}
            \For{$\varepsilon$ in $\epsilon$}
            \For{$l$ in $L$}
            \For{$v$ in $\{ N \times l \}$}
            \For{$\overline{v}$ in $\{ (N \times l) - v \}$}
            \If{$\tau_{v}^{l} \neq \tau_{\overline{v}}^{l}$}
            \If{$BernoulliTrial(p=w_{\dot{\tau_{n}}}^{\tau_{n} - \tau_{v}^{l} + \tau_{\overline{v}}^{l}}) = 1$}
            \State $\tau_{v}^{l} := \tau_{\overline{v}}^{l}$
            \State \textbf{go to } \texttt{line 4}
            \EndIf
            \EndIf
            \EndFor
            \EndFor
            \EndFor
            \EndFor
        \EndProcedure
    \end{algorithmic}
\end{algorithm}

\section{\textit{Network Diffusion} Package}

Keeping in mind the way we defined the multiprocess interaction framework, it is possible to move on to the package description. We also find it useful to present some high-level aspects of its architecture.

The most important functional requirements were paired with the questions posed in the introduction of sec.~\ref{sec:operating_principle}, i.e. the possibility (1)~to simulate any number of processes during the experiment and (2)~to define interactions between them. We also wanted to support experiments on external data (the commonly used format to store multilayer networks - \textit{.mpx} has been chosen) and generate data and figures reflecting changes of states in the network during experiments.

When it comes to nonfunctional requirements, the main was the programming language. Due to the popularity of particular ones in data science, we faced an alternative: Python or R. The popularity and support of both languages are very similar, so in the end, the first one was chosen subjectively. It is worth noting that we were not focused on efficiency. Therefore, the implementation was reflecting alg.~\ref{simalgo} without optimisation.

\subsection{Implementation}

The whole package is based on the NetworkX library since, as shown in sec.~\ref{sec:other_tools}, it has a lot of built-in functionalities, is widely used, and has great community support. This significantly reduces the so-called entry threshold of the package for new users. Following the theoretical concepts, we divide the code into three main components: \textit{MultilayerNetwork}, \textit{PropagationModel}, \textit{MultiSpreading}.

The \textit{MultilayerNetwork} was able to store multilayer networks (as a dictionary of \textit{networkx.Graph} objects keyed by the layer names). We also added integrated I/O methods with the NetworkX library for \textit{.mpx} files.

As for the \textit{PropagationModel} class, the biggest problem to solve was to define the model unambiguously according to def.~\ref{def:metamodel}. We solved it by defining the following procedure of propagation model initialisation: (1)~user passes processes to the model in the form of name and internal states for each process, (2)~user inserts the default weight for transition, (3)~user calls a compilation function (then the model is converted to an interpretable form for the simulation environment), (4)~user manually changes the weights of particular transitions (usually those which are significant for the experiment). Regarding the problem of defining a data structure to keep the model in, during compilation, a unique graph with only possible transitions has been created for each process. This made easy referring to particular transitions in the model by operating strings of type \textit{<state>.<process>}, similar to the SimInf library~\ref{subsec:close_competition}.

\begin{figure*}
	\centering
	\includegraphics[width=0.7\linewidth]{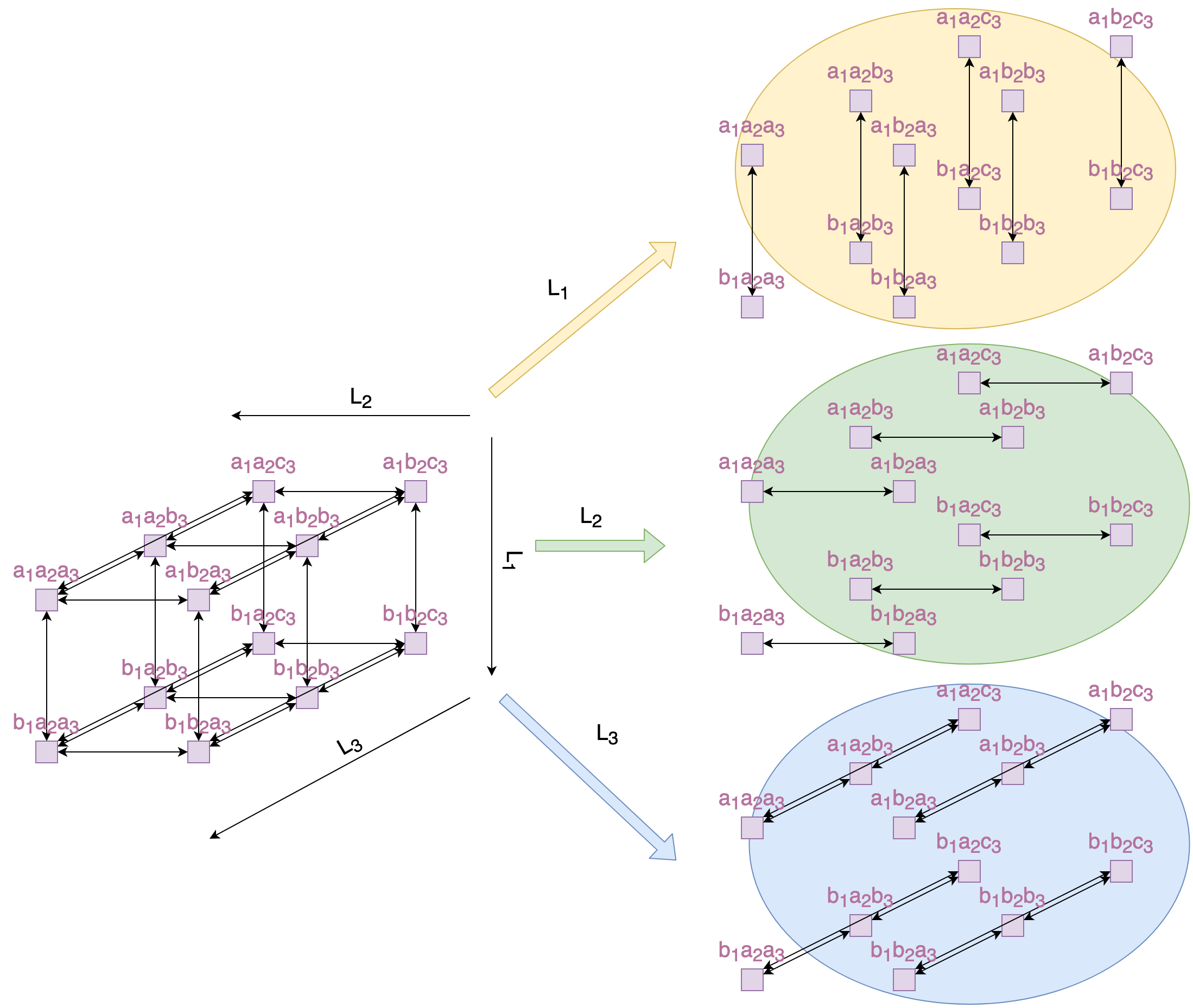}
	\caption[]{Scheme for compiling a propagation model. A single, three-dimensional model is stored as a structure of three separate NetworkX graphs, one for each process, with edges covering allowed only transitions.}
\end{figure*}

The \textit{MultiSpreading} was compacting the modules described already into a coherent whole. It had two fields \textit{\_network} and \textit{\_model}, which were associated with each other by process names (i.e. for each process in the model, there had to be a corresponding layer in the network with the same name). The other relevant methods were \textit{set\_initial\_states} and \textit{perform\_propagation}. The first one was used to set the initial conditions of the simulation - what percentage/number of nodes is currently in which state? The second method, on the other hand, was responsible for the simulation itself, according to the alg.~\ref{simalgo}. To separate the preprocessing of the experiment and the results from the simulation, a class \textit{ExperimentLogger} has been prepared, whose object is returned after a successful experiment run.

\begin{table}[ht]
    \renewcommand{\arraystretch}{1.3}
    \caption[]{Artefacts related to the \textit{Network Diffusion} package.}
    \label{tab:artefacts}
    \centering
    \begin{tabular}{|p{0.2\columnwidth}|p{0.35\columnwidth}|p{0.31\columnwidth}|}
    \hline
    \textbf{Artefact} & \textbf{Link} & \textbf{Description} \\ \hline \hline
    Main repository & \url{https://github.com/anty-filidor/network_diffusion} 
    & Main repository with package's code~\cite{czuba2021zenodo} 
    \\ \hline
    PyPi package & \url{https://pypi.org/project/network-diffusion/}
    & Built package ready to download \\ \hline
    Anaconda package & \url{https://anaconda.org/anty-filidor/network_diffusion} 
    & Built package ready to download \\ \hline
    Documentation & \url{https://network-diffusion.readthedocs.io} 
    & Web page with comprehensive documentation \\ \hline
    Examples (source code) & \url{https://github.com/anty-filidor/network_diffusion_examples} 
    & Auxiliary repository with examples\\ \hline
    Runnable capsule with examples & \url{https://codeocean.com/capsule/8807709/tree/v4} 
    & Examples ready to run online via web browser~\cite{czuba2021codeocean}\\ \hline
    \end{tabular}
\end{table}

\section{Example of usage}\label{sec:example}

Finally, a practical example can be presented. We prepared three comprehensive simulation scripts for this purpose: (1)~example of epidemic propagation combined with vaccinations and awareness of the population, (2)~marketing campaign of two competitive products, (3)~gossip spreading on two different social networks. All are included in the GitHub repository~\cite{czuba2021zenodo} and the Code Ocean capsule~\cite{czuba2021codeocean}. In this section, we present the first one (i.e. \textit{epidemic.py}) in shortened form\footnote{It is worth noting that the remaining two examples utilise other mechanisms of phenomena spreading. They are all in the form of the Jupyter Notebook, and we welcome readers to investigate them.}. To avoid coping scripts, we will refer the reader to the corresponding lines of code in the repository/capsule.

\subsection{Scenario to simulate}

In this experiment, we will simulate the spreading of three processes: an illness affecting the network, awareness of its existence, and vaccination against it. Keeping in mind the notation introduced in tab.~\ref{domain_dict} we will first define a propagation model:

\begin{itemize}
    \item $illness: T = [s, i, r], W = [w_{s}^{i}, w_{i}^{r}]$, where $s$ denotes suspected node, $i$ infected and $r$ recovered,
    \item $awareness: T = [n, a], W = [w_{n}^{a}]$, where $n$ means not-aware node and $a$ respectively aware one,
    \item $vaccination: T = [u, v], W = [w_{u}^{v}]$, where $u$ means not vaccinated node and $v$ vaccinated one.
\end{itemize}

The compiled model can be represented by a three-dimensional grid with twelve points. Regarding the fact that we defined only a subset of all possible transitions (e.g., $w_{i}^{s}$ was discarded), we have to define "only" twenty ones:

\begin{itemize}
    \item for states varying in the $illness$: $w_{snu}^{inu}$, $w_{inu}^{rnu}$, $w_{sau}^{iau}$, $w_{iau}^{rau}$, $w_{snv}^{inv}$, $w_{inv}^{rnv}$, $w_{sav}^{iav}$, $w_{iav}^{rav}$,  
    \item for states varying in the $awareness$: $w_{snu}^{sau}$, $w_{inu}^{iau}$, $w_{rnu}^{rau}$, $w_{snv}^{sav}$, $w_{inv}^{iav}$, $w_{rnv}^{rav}$, 
    \item for states varying in the $vaccination$: $w_{sau}^{sav}$, $w_{iau}^{iav}$, $w_{rau}^{rav}$, $w_{snu}^{snv}$, $w_{inu}^{inv}$, $w_{rnu}^{rnv}$, 
\end{itemize}

Let us also add more constraints to the experiment: we want both vaccination and awareness to reduce the probability of getting sick, and awareness of the disease should make the nodes more willing to get vaccinated. We can obtain such an effect by setting some of the transitions listed above to predefined values. Moreover, we note that some of them seem illogical. For example, $w_{snu}^{snv}$ one, because it is hard to imagine that an individual can be vaccinated without even being aware of the disease. Hence, in our model, we can limit the transitions to the following set of non-zero ones:

\begin{itemize}
    \item for states varying in $illness$: $w_{snu}^{inu}=0.4$, $w_{inu}^{rnu}=0.1$, $w_{sau}^{iau}=0.2$, $w_{iau}^{rau}=0.3$, $w_{sav}^{iav}=0.05$, $w_{iav}^{rav}=0.7$,  
    \item for states varying in $awareness$: $w_{snu}^{sau}=0.05$, $w_{inu}^{iau}=0.2$, $w_{snv}^{sav}=1$,
    \item for states varying in $vaccination$: $w_{sau}^{sav}=0.03$, $w_{iau}^{iav}=0.1$.
\end{itemize}

On the other hand, in rare cases, situations like $snu \xrightarrow{} snv$ can be possible. Therefore, we will set up some tiny weights ($0.005$) for all possible but unlikely to happen transitions.

\subsection{Defining a propagation model}

With these assumptions, we can finally prepare the script. As a first step, we have to import all required packages (see lines 2:16). After that, we can step forward to define a propagation model (see lines 29:48). Please pay attention to the function \pyth{compile}. When the keyword argument is set, all theoretically possible transitions in the model will have a weight assigned to \pyth{0.005}. With this operation, we can increase the model's flexibility to include some edge cases.

\subsection{Defining a network}

The next step is to define the network. As mentioned in sec.~\ref{subsec:sim_framework}, it should have as many layers as the processes spreading within it. We can achieve this by duplicating a flat NetworkX graph into three layers. In this experiment, we will use a predefined graph reflecting interactions in “Les Misérables” (see lines 50:52).

\subsection{Defining and running an experiment}

After that, we can define and start the experiment (see lines 54:99). We set the initial state of the \pyth{"ill"} process to have \pyth{65} nodes \pyth{"s"}, \pyth{10} \pyth{"i"} and \pyth{2} \pyth{"r"} at the first epoch. Here, we did it manually for particular nodes to obtain reproducibility, but \textit{Network Diffusion} contains another function that does it for randomly selected nodes at once: \pyth{nx.Experiment.set_initial_states}.

\subsection{Results and discussion}

\begin{figure}[ht]
    \centering
    \subfloat[]{\includegraphics[width=0.87\columnwidth]{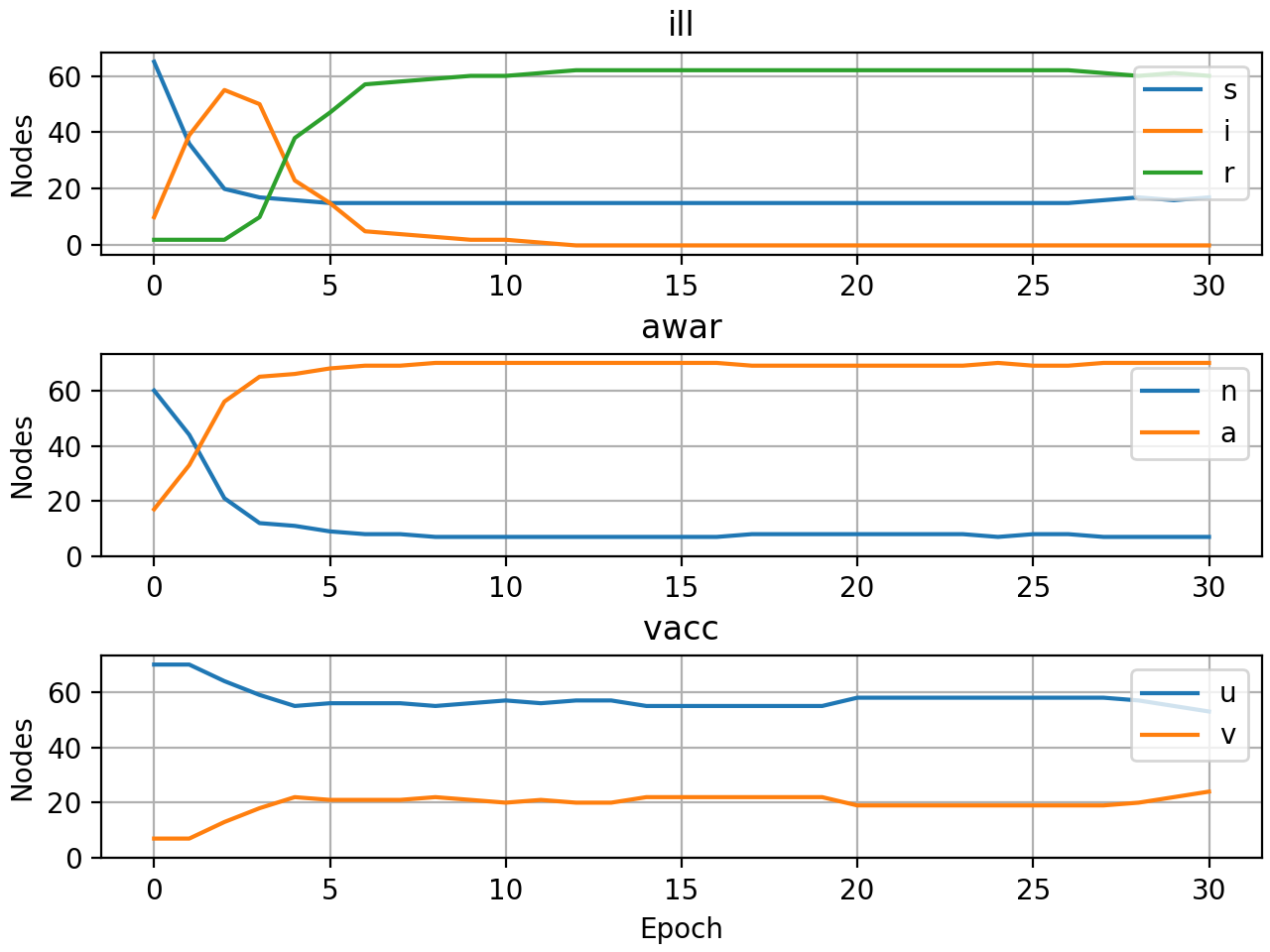}\label{fig:ex1_bulk}}\quad
    \subfloat[]{\includegraphics[width=0.87\columnwidth]{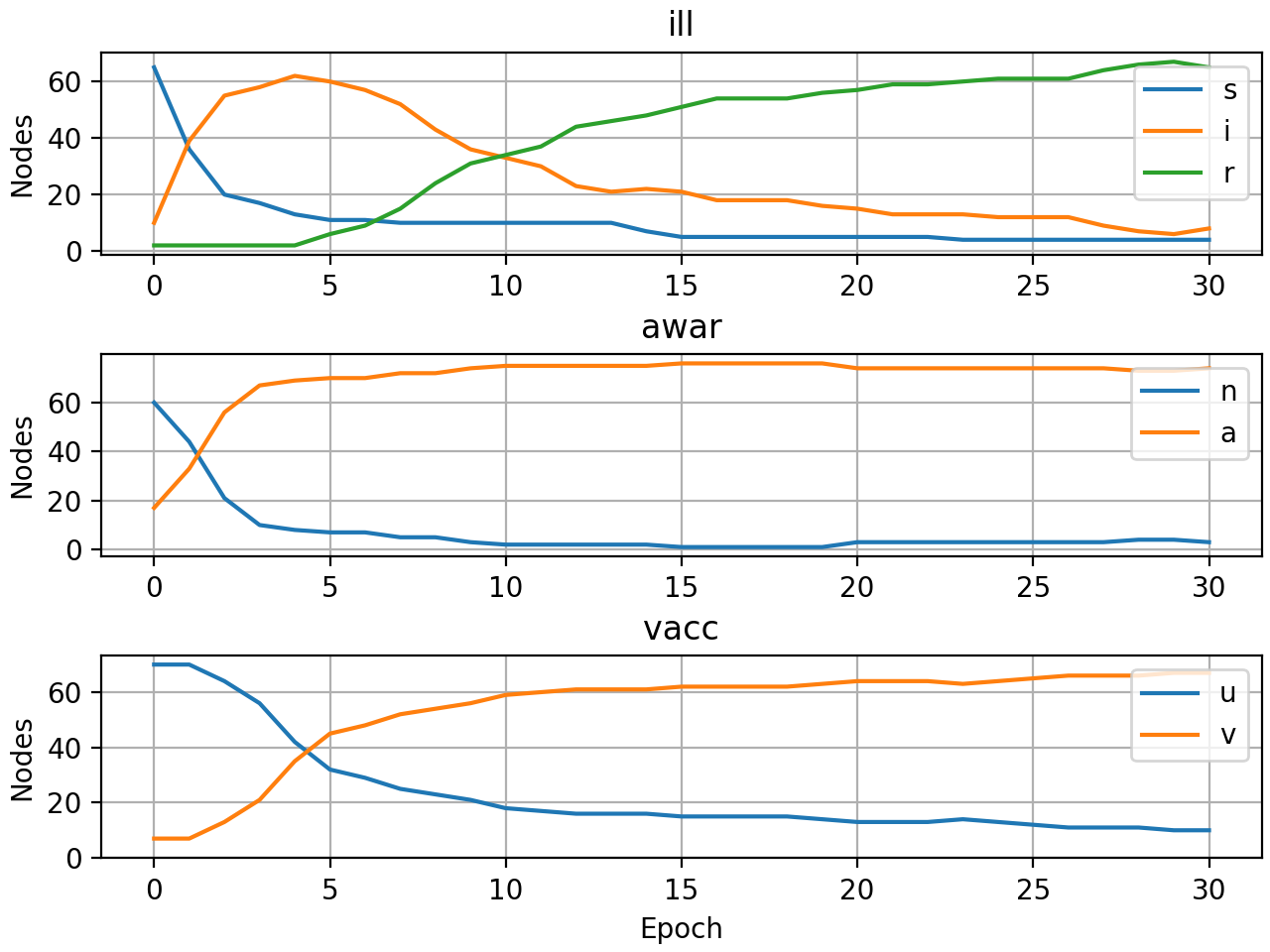}\label{fig:ex2_bulk}}
    \caption[]{Visualisations of experiments - dynamics of processes in the network (a) according to presented assumptions, (b) with modified \textit{illness} to make it more contagious.}
\end{figure}

Once the simulation is completed, an \pyth{ExperimentLogger} object is returned. We can use it to prepare a report (see line 99). As a result, we obtain (snippets available in documentation - see tab.~\ref{tab:artefacts}): propagation report for each process as set of \textit{.csv} files containing information about the distributions of particular states in each epoch (\textit{awar\_propagation.csv}, \textit{ill\_propagation.csv}, \textit{vacc\_propagation.csv}), information about the network used for experiment (\textit{network\_report.txt}), the model (\textit{model\_report.txt}) and visualisation of the experiment (fig.~\ref{fig:ex1_bulk}).

As we can see in fig.~\ref{fig:ex1_bulk}, interactions between processes reflect the assumed scenario, the more individuals vaccinated and aware, the less increase of ill nodes in the network. 
On the other hand, in fig.~\ref{fig:ex2_bulk}, we can see a variation of the experiment. In order to obtain the effect of more malicious illnesses that cannot be so easily mitigated with vaccines and awareness of the population, we have modified weights $w_{inu}^{rnu}$, $w_{iau}^{rau}$ and $w_{iav}^{rav}$. That resulted in a slower falling curve for state \textit{i} that allowed the $vaccination$ process to continue spreading among the network.

\section{Conclusions}

In this paper, we have introduced a computational approach to simulate the spreading of multiple processes in networks. 
Keeping in mind the analysis of the existing tools from sec.~\ref{sec:competition}, we find it more valuable since it fulfils the observed gap in the domain. Despite the basic character of \textit{Network Diffusion}, we hope that it will help other researchers in analysing and understanding spreading processes in networks.

We would like to mention that there is great potential in the further development of the package. As we described in sec.~\ref{sec:operating_principle}, for now, we are limited to a fairly simple schema of the experiment. We plan to extend it in the next releases by aspects of time dynamics, such as adding layer switching cost for the spreading process~\cite{min2016layer}. We also noticed a need to provide the user with the possibility to customise a function that changes the states of nodes. Currently, we have a fairly simple mechanism that depends on iteration through neighbours combined with the Bernoulli test. However, we can use another approach, e.g. that takes into account all neighbours of the node at once. Another important aspect is an enhancement of the package's computational capabilities (see sec.~\ref{apdx:computional_efficiency} for details). Finally, the concept that is currently investigated by the authors is a development of a training mechanism that sets up weights of the models in order to fit with real data, which allows for predicting further spreading of the phenomena. 

Having written all that, we can say that \textit{Network Diffusion} 
may help to address the problem of multiple spreading processes within the networks. This issue is certainly real, and the COVID-19 crisis shows that we need to be able to simulate such phenomena by taking into account coexisting phenomena. As we demonstrated in the article, the package has a fine theoretical background and examples proving its usability. We strongly believe that our concept can be helpful and become an inspiration to other projects.

\bibliographystyle{IEEEtran} 
\bibliography{references}

\appendix

\subsection{Computational efficiency of the package}\label{apdx:computional_efficiency}

\begin{figure}[ht]
    \centering
    \includegraphics[width=0.97\columnwidth]{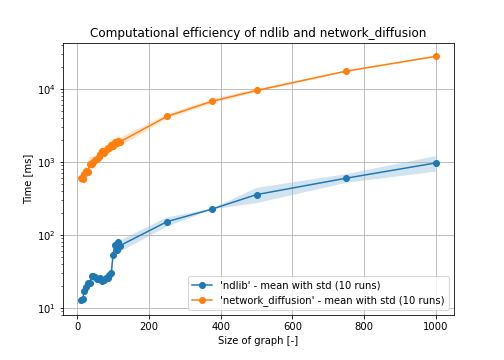}
    \caption[]{Comparison of the efficiency of the Network Diffusion and the NDlib.}
    \label{fig:effic}
\end{figure}

Since \textit{Network Diffusion} has been considered a prototype framework, we have decided to investigate its time efficiency compared to NDlib~\cite{ndlib}. 
We performed an experiment on a single-layer SIR model (due to limitations of the NDLib, we could not test a more complex case) spreading within the Erdos-Renyi graph, the size of which varied. For each network size, both implementations were executed 10 times (to mitigate the side effects that burden the processor). Each call to the function was measured in milliseconds. The results are presented in fig.~\ref{fig:effic}. The time efficiency of Network Diffusion is a field for improvement, especially for large networks.

\subsection{Running attached code examples}

There are two ways to run the examples: (1) local execution, (2) via the Code Ocean capsule~\cite{czuba2021codeocean}. Despite the differences in code handling, both are expected to produce the same results. To follow the first, the user needs to have Python installed. It is recommended to create a virtual environment based on the \textit{requirements.txt} file. Then, to run the chosen example, the corresponding script needs to be executed. The second method requires far less effort. To run our code, we recommend first that you duplicate the capsule into a self-owned account. Then, to run the engine, the "reproducible run" button has to be pressed. To run a Jupyter notebook, the user has to select a Jupyter icon from the list below the mentioned button.

\end{document}